\documentclass[12pt]{article}
\usepackage{amssymb,amsmath,cite,geometry}
\catcode`\@=11

\@addtoreset{equation}{section}
\def\theequation{\arabic{section}.\arabic{equation}}
     
\catcode`\@=11
\def\thesection{\arabic{section}}

\def\appendix{\setcounter{section}{0}
        \def\thesection{Appendix.}
        \def\theequation{\Alph{section}.\arabic{equation}}}
\def\section{\@startsection{section}{1}{\z@}{3.5ex plus 1ex minus
   .2ex}{2.3ex plus .2ex}{\large\bf}}
\def\subsection{\@startsection{subsection}{1}{\z@}{3.5ex plus 1ex minus
   .2ex}{2.3ex plus .2ex}{\bf}}
     
\long\def\@makefntext#1{\parindent 0cm\noindent
\hbox to 1em{\hss$^{\@thefnmark}$}#1}

\newcommand{\captionfonts}{\small}
\makeatletter  
\long\def\@makecaption#1#2{%
  \vskip\abovecaptionskip
  \sbox\@tempboxa{{\captionfonts #1: #2}}%
  \ifdim \wd\@tempboxa >\hsize
    {\captionfonts #1: #2\par}
  \else
    \hbox to\hsize{\hfil\box\@tempboxa\hfil}%
  \fi
  \vskip\belowcaptionskip}
\makeatother   

\begin{document}
\begin{titlepage}
\vspace{.5in}
\begin{flushright}
July 2011\\  
\end{flushright}
\vspace{.5in}
\begin{center}
{\Large\bf
 Effective Conformal Descriptions\\[.4ex] of Black Hole Entropy}\\  
\vspace{.4in}
{S.~C{\sc arlip}\footnote{\it email: carlip@physics.ucdavis.edu}\\
       {\small\it Department of Physics}\\
       {\small\it University of California}\\
       {\small\it Davis, CA 95616}\\{\small\it USA}}
\end{center}

\vspace{.5in}
\begin{center}
{\large\bf Abstract}
\end{center}
\begin{center}
\begin{minipage}{4.75in}
{\small
It is no longer considered surprising that black holes have temperatures 
and entropies.  What remains surprising, though, is the universality of 
these thermodynamic properties: their exceptionally simple and general 
form, and the fact that they can be derived from many very different 
descriptions of the underlying microscopic degrees of freedom.  I review 
the proposal that this universality arises from an approximate conformal 
symmetry, which permits an effective ``conformal dual'' description that 
is largely independent of the microscopic details.
}
\end{minipage}
\end{center}
\end{titlepage}
\addtocounter{footnote}{-1}

\section{Introduction}

Since the seminal work of Bekenstein \cite{Bekenstein} and Hawking 
\cite{Hawking}, we have understood that black holes behave as
thermodynamic systems, with characteristic temperatures and entropies.  
A central task for quantum gravity is finding a statistical mechanical 
description of this behavior in terms of microscopic states.  One 
important clue may come from the surprising ``universality'' of these 
thermodynamic properties, which manifests itself in two ways:
\begin{itemize}
\item[--]The Bekenstein-Hawking entropy
\begin{align}
S = \frac{\mathcal{A}}{4\hbar G}
\label{a1}
\end{align} 
takes the same simple form for all black holes, independent of the
charges, spin, and even the number of spacetime dimensions.  If the 
action differs from that of general relativity, the entropy is modified, 
but again in a simple and universal way \cite{Wald}. 
\item[--]While the ``correct'' microscopic origin of black hole 
entropy is not settled, we now have a number of candidates, 
coming from string theory, loop quantum gravity, holographic 
entanglement entropy, Sakharov-style induced gravity, and 
several other approaches.\footnote{See \cite{Carlip} for a 
more detailed discussion.}   But although these methods count 
very different states---or, in the case of Hawking's original derivation
\cite{Hawking}, seem to know nothing about quantum gravitational 
states at all---they all yield the same entropy.  To be sure, no approach 
is yet complete, and each requires additional assumptions.  But within 
its domain of validity, each seems to work fairly well, and the 
universal agreement with (\ref{a1}) remains deeply mysterious.
\end{itemize}

One attempt to explain this universality is to posit that black hole
thermodynamics has an effective ``dual'' description in terms of a 
two-dimensional conformal theory, whose key properties are  
governed by symmetries.  Black holes are, of course, neither 
two-dimensional nor conformally invariant, so at first sight this 
proposal seems rather surprising.  But in the sixteen years or so 
since such a description was first proposed \cite{Carlipa}, the idea
has become considerably more plausible.  In special cases---in
particular, the (2+1)-dimensional BTZ black hole \cite{BTZ}---the 
two-dimensional conformal description seems quite convincing 
\cite{Strominger,BSS,Carlipb}, although subtle questions about the 
conformal field theory remain \cite{Witten,Maloney}.  For more 
general black holes \cite{Carlipc,Solodukhin,Carlipd}, the issue 
is much less settled, but the recent discovery of an extremal 
Kerr/CFT correspondence \cite{Guica} has brought a good deal
of renewed interest.

In this paper, I will review the arguments for a two-dimensional conformal 
description---a CFT dual, in the current string-theory-inspired language%
---for a variety of black holes.  I will start by sketching the general
derivation of such a dual; then describe a few examples; and finally 
suggest a physical explanation for the Bekenstein-Hawking entropy.  My 
purpose is not to explain any specific instance of this construction in depth; 
for that, the reader should refer to the primary literature.  Rather,
I will try to describe the general framework in which to think about CFT
duals of black holes.

\section{Building a black hole/CFT correspondence in four easy steps
\label{secSteps}}

While the details differ, most derivations of black hole/CFT duality 
are based on the same four steps:
\begin{enumerate} 
\item[i.] Find an appropriate boundary and impose boundary conditions that
specify properties of the black hole;
\item[ii.] Determine how these boundary conditions affect the symmetries of
general relativity, the algebra of diffeomorphisms or ``surface deformations'';
\item[iii.] Look for a preferred subalgebra of diffeomorphisms of the circle
($\operatorname{Diff}S^1$), or perhaps two copies 
($\operatorname{Diff}S^1\times\operatorname{Diff}S^1$);
\item[iv.] Use standard methods from conformal field theory---in particular,
the Cardy formula \cite{Cardya,Cardyb}---to extract physical information
from this symmetry.
\end{enumerate}
I will explain each of these in turn.

\subsection{Find an appropriate boundary and impose boundary conditions}

A question about a quantum black hole is almost always a question 
about \emph{conditional} probabilities: ``If a black hole with certain 
characteristics is present, what is the probability that \dots?''  One must 
therefore decide how to impose the conditions, that is, the presence and 
specific characteristics of the black hole.  Although there has been some 
work on adding these conditions as explicit constraints \cite{Carlipe,Carlipf}, 
probably the easiest approach is probably to impose them as boundary 
conditions.

This is most easily done for the (2+1)-dimensional black hole of 
Ba{\~n}ados, Teitelboim, and Zanelli (BTZ) \cite{BTZ}.  This black hole
lives in an asymptotically anti-de Sitter spacetime, and as Brown and
Henneaux showed \cite{BrownH}, the asymptotic boundary conditions at 
infinity determine its mass and spin.  Many near-extremal black holes 
in more than three spacetime dimensions have a near-horizon region 
that closely resembles (2+1)-dimensional anti-de Sitter space 
\cite{Strominger,Skenderis}, and Brown-Henneaux boundary conditions 
can be imposed at the ``boundary'' of this region.  Similarly, extremal 
Kerr black holes have a near-horizon region that looks like anti-de 
Sitter space \cite{BardeenH}, and one can impose modified AdS-like 
boundary conditions there \cite{Guica}.

An alternative, and perhaps more intuitive, choice is to treat the
event horizon itself as a boundary \cite{Carlipc,Solodukhin,Carlipd}.
The horizon is not, of course, a real boundary: an infalling observer 
can pass through it freely.  It is, however, a place at which one can 
restrict the metric to impose the condition that a particular type of 
black hole be present.  

Such a restriction is essentially a boundary condition.  In the path 
integral formalism, for instance, it amounts to splitting spacetime 
into two pieces along the horizon, with separate path integrals for each 
piece, each with a suitable boundary condition.  For two-dimensional 
free field theories \cite{CarlipDLP} and three-dimensional Chern-Simons 
theory \cite{WittenCS}, such splittings have been investigated carefully,
and it has been shown that one can reconstruct the full path integral 
by ``gluing'' the two halves and integrating over boundary data.  This
last integration can be viewed as a quantum mechanical sum over a 
complete set of intermediate states; the role of the boundary conditions 
is to determine which set of intermediate states appears.

For quantum gravity, the imposition of boundary conditions at the
true horizon is difficult, essentially because the horizon is a null
surface.  In practice, one must instead impose boundary conditions 
on a ``stretched horizon'' \cite{Thorne}, a timelike surface just outside 
the horizon, and then take a limit as the surface shrinks to the horizon.  
This is not an entirely satisfactory approach, though---the results 
sometimes depend on the limiting procedure \cite{Carlipg}---and 
there is clearly room for improvement.

\subsection{Determine how these boundary conditions affect the 
symmetries}

In its Hamiltonian form, general relativity in $n$ spacetime dimensions 
is a constrained theory, with $n$ first-class constraints $\mathcal{H}_\mu$
\cite{Kiefer}.  These constraints generate the symmetries of the theory.  
In the covariant phase space formalism \cite{Walda,Ashtekar,Crnkovic}, 
these symmetries are the spacetime diffeomorphisms.  That is, for 
a diffeomorphism generated by a vector field $\xi^\mu$, the generators
\begin{align}
H[\xi] = \int_\Sigma d^{n-1}x\,\xi^\mu\mathcal{H}_\mu ,
\label{b1}
\end{align}
integrated over a Cauchy surface $\Sigma$, have Poisson brackets
\begin{align}
\left\{ H[\xi], F\right\} = \delta_\xi F 
\label{b2}
\end{align}
for any phase space function $F$, where $\delta_\xi F$ is the variation
of $F$ under the diffeomorphism generated by $\xi$.  Further, these 
generators obey an algebra
\begin{align}
\left\{ H[\xi], H[\eta]\right\} = H[\{\xi,\eta\}]  ,
\label{b3}
\end{align}
where $\{\xi,\eta\}$ is the ordinary commutator of vector fields.

The ADM formalism \cite{ADM} requires some minor modifications: the 
parameters $\xi^\mu$ are replaced by ``surface deformation parameters'' 
$\xi^\perp$ and ${\hat\xi}^i$ \cite{Teitelboim}, given by
\begin{align}
\xi^\perp = N\xi^t, \qquad {\hat\xi}^i = \xi^i + N^i\xi^t ,
\label{b4}
\end{align}
and the brackets in (\ref{b3}) become the surface deformation brackets
\begin{align}
&\{\xi,\eta\}_{\scriptscriptstyle\mathit{SD}}^\perp 
      = {\hat\xi}^iD_i\eta^\perp - {\hat\eta}^iD_i\xi^\perp \nonumber\\
&\{\xi,\eta\}_{\scriptscriptstyle\mathit{SD}}^i
      = {\hat\xi}^kD_k{\hat\eta}^i - {\hat\eta}^kD_k{\hat\xi}^i
      + q^{ik}\left(\xi^\perp D_k\eta^\perp - \eta^\perp D_k\xi^\perp\right) ,
\label{b5}
\end{align}
where $q_{ij}$ is the spatial metric on $\Sigma$ and $D_i$ is the
corresponding covariant derivative.  On shell, however, the two formalisms 
are essentially the same \cite{Ryan}.

For a manifold with boundary, the story becomes a bit more complicated.  
As Regge and Teitelboim first noted \cite{ReggeT}, if $\Sigma$ has a
boundary, the generators (\ref{b1}) are not ``differentiable''---a 
variation of the fields gives not only the usual functional derivative
of $\mathcal{H}_\mu$, but also an added boundary term arising from
integration by parts.  As a result, the Poisson brackets (\ref{b2}) and
(\ref{b3}) acquire new, singular boundary contributions.  

While it may be possible to generalize the Poisson algebra to include 
such terms \cite{Soloviev}, the simplest remedy is to add a boundary
term to the generators,
\begin{align}
{\bar H}[\xi] = H[\xi] + B[\xi]
    = \int_\Sigma d^{n-1}x\,\xi^\mu\mathcal{H}_\mu 
    + \int_{\partial\Sigma} d^{n-2}x\,\xi^\mu\mathcal{B}_\mu ,
\label{b6}
\end{align}
where the boundary term $B[\xi]$ is chosen to cancel the boundary 
terms coming from the partial integration.  The particular form of $B[\xi]$ 
depends on the choice of boundary conditions, since these determine 
what is \emph{not} varied at $\partial\Sigma$.  Note that while
$H[\xi]$ vanishes on shell, ${\bar H}[\xi]$ need not.  It---or, 
equivalently, the boundary term $B[\xi]$---may be viewed as a Noether
charge associated with the corresponding symmetry \cite{Walda}.
In particular, for a boundary at infinity, suitable choices of $\xi$ 
give the mass and angular momentum.

In general, the new boundary term may alter the Poisson brackets
(\ref{b3}), which become
\begin{align}
\left\{ {\bar H}[\xi], {\bar H}[\eta] \right\} = {\bar H}[\{\xi,\eta\}]  
    + K[\xi,\eta] ,
\label{b7}
\end{align}
where $K[\xi,\eta]$ is a central term.  This term was first computed by
Brown and Henneaux for (2+1)-dimensional asymptotically anti-de Sitter
space \cite{BrownH}, where it takes a simple form that depends only
on Newton's constant and the cosmological constant.  A more general
form, again computed in the ADM formalism, will be given below.

An alternative approach the the computation of this central term is
based on the covariant canonical formalism \cite{Walda,Ashtekar,Crnkovic}.  
This approach starts with the observation that in a well-posed theory,
the phase space, considered as the space of initial data, is isomorphic
to the space of classical solutions.  One can therefore carry out a 
canonical analysis on this space of solutions, in a covariant manner
that does not require a splitting of spacetime into space and time.  The
covariant canonical derivation of the central term $K[\xi,\eta]$ was first
treated in \cite{Carlipd}; the results were corrected and generalized 
in \cite{Koga,Silva}, and greatly extended in \cite{Barnich,Compere}.

\subsection{Look for a preferred subalgebra of diffeomorphisms of the circle}

The central term $K[\xi,\eta]$ in the algebra (\ref{b7}) depends strongly 
on the boundary conditions.  Suppose, however, that one can find a 
subalgebra $\operatorname{Diff}S^1$ of diffeomorphisms of the circle, that is,
a one-parameter subalgebra such that
\begin{align}
\{\xi,\eta\} = \xi\eta' - \eta\xi' .
\label{c1}
\end{align}
One can then use the general result that the such an algebra has an almost
unique central extension \cite{GelfandF}: up to redefinitions of the 
generators,
\begin{align}
\left\{ {\bar H}[\xi], {\bar H}[\eta] \right\} = {\bar H}[\{\xi,\eta\}]  
    +  \frac{c}{48\pi}\int d\varphi\, (\xi'\eta'' - \eta'\xi'') .
\label{c2}
\end{align}
The algebra (\ref{c2}) is known as a Virasoro algebra; the parameter $c$
is the central charge.

It is at this point that two-dimensionality and conformal symmetry enter
the picture.  The Virasoro algebra, viewed as the algebra of holomorphic
diffeomorphisms of a two-manifold, is the fundamental symmetry algebra 
of a two-dimensional chiral conformal field theory \cite{Francesco}.  
Similarly, $\operatorname{Diff}S^1\times\operatorname{Diff}S^1$, viewed 
as an algebra of holomorphic and antiholomorphic diffeomorphisms, is
the symmetry algebra of a nonchiral conformal field theory.

The existence of a preferred $\operatorname{Diff}S^1$ in the higher-%
dimensional algebra (\ref{b7}) is certainly not guaranteed.  But it is also 
not unreasonable to expect such a subgroup.
The BTZ black hole, for instance, is a three-dimensional asymptotically
anti-de Sitter spacetime.  The boundary of $\operatorname{AdS}_3$
is a flat two-dimensional cylinder, so the existence of a pair of Virasoro
algebras in the asymptotic symmetry group is no surprise.  Similarly,
the near-horizon extremal Kerr metric of \cite{BardeenH} contains a
twisted $\operatorname{AdS}_3$ factor, so a $\operatorname{Diff}S^1$
subgroup is again not too surprising.

For more general black holes, the existence of a $\operatorname{Diff}S^1$
subgroup is less obvious.  But generic black holes \emph{do} have preferred
directions.  A stationary black hole in 3+1 dimensions admits two Killing
vectors, for instance, representing time translations and rotations; in 
the extremal \cite{Guica} and near-extremal \cite{Rasmussen} Kerr/CFT
correspondences, these pick out the relevant $\operatorname{Diff}S^1$
factors.  

Even more generally, for any ``equilibrium'' black hole---any black hole 
characterized by an isolated horizon \cite{Ashtekarb} with a stationary 
neighborhood---the horizon is a Killing horizon \cite{Date}.  This means
the horizon $\mathcal{H}$ is a null surface generated by a local Killing 
vector $\chi^a$ that becomes null at $\mathcal{H}$.  This Killing vector
determines a one-dimensional subalgebra of diffeomorphisms for a 
generic black hole, giving the  $\operatorname{Diff}S^1$ factor used 
in \cite{Carlipd} and \cite{Carlipg}.  The same Killing vector is important 
elsewhere in black hole thermodynamics: it fixes the temperature of fields 
in a black hole background \cite{Carlip}, and defines the basic symmetry 
required for the proof of the generalized second law of thermodynamics
 \cite{Wall}.

\subsection{Use conformal field theory methods to extract physical 
information}

The symmetries of a theory give us a powerful tool for extracting physical 
information.  This is particularly true for a two-dimensional conformal field 
theory, for which the group of symmetries is infinite-dimensional.  In fact, the
conformal symmetries largely determine such key features as correlation 
functions \cite{Francesco}.  Even more important for our purposes, they
determine the asymptotic density of states, that is, the entropy.

Consider a conformal field theory whose holomorphic symmetries are 
given by the Virasoro algebra (\ref{c2}).  As noted earlier, the zero-mode 
${\bar H}[\xi_0]$ is a kind of charge, sometimes called the conformal
dimension or the conformal weight.  Suppose the spectrum of this operator 
is nonnegative, with a lowest value $\Delta_0$.  Then the asymptotic 
density of states at a fixed value $\Delta$ of the weight is 
\cite{Cardya,Cardyb}
\begin{align}
\ln \rho \sim 2\pi
   \sqrt{\frac{c_{\hbox{\tiny\it eff}}}{6}\left(\Delta - \frac{c}{24}\right)}
\qquad \hbox{with}\ c_{\hbox{\tiny\it eff}} = c-24\Delta_0  .
\label{d1}
\end{align}
This remarkable result, known as the Cardy formula, determines the
microcanonical entropy in terms of a few parameters that characterize
the symmetry, independent of any other details of the theory.   Logarithmic 
\cite{Carlipi} and higher order \cite{Birmingham,Loran} corrections 
are known as well.  The derivation of the Cardy formula is fairly 
straightforward---see, for instance, \cite{Carliph}---but it involves a
duality between high- and low-energy states; I think it is fair to say that 
we do not yet have a very deep physical understanding of the result. 

The Cardy formula comes in a canonical version as well.  Instead of
fixing the weight $\Delta$, let us fix the temperature $T$.  Then
(see, for instance, section 8 of \cite{Bousso})
\begin{align}
S\sim \frac{\pi^2}{3}cT  .
\label{d2}
\end{align}
There is a slightly subtle point here: temperature is normally associated
with periodicity in imaginary time, but here the relevant periodicity is
that of the $S^1$ picked out by the Virasoro algebra.  We will see
examples of this in section \ref{seceg}.

Going beyond the entropy, one can also use conformal field theory 
techniques to explore such matters as greybody factors, superradiance, 
and the Hawking temperature (see, for example, 
\cite{Maldacena,Bredberg,Emparan}).  For the most part, however, 
such work lies outside the scope of this review.

\section{Central terms in the ADM formalism \label{secADM}}

The discussion in the preceding section was somewhat abstract, and it
is useful to look at some specific examples.  To do so, I must first fill in 
a few more details.  I will work in the ADM approach, which is perhaps less 
elegant than the covariant phase space method, but also requires less 
formalism.

In $n$ spacetime dimensions, the ADM form of a general metric is
\begin{align}
ds^2 = -N^2dt^2 + q_{ij}(dx^i+N^idt)(dx^j+N^jdt)  , \quad 
    (i = 1,2,\dots,n-1) .
\label{e0}
\end{align}
The momentum conjugate to the spatial metric $q_{ij}$ is
\begin{align}
\pi^{ij} =  \sqrt{q}(K^{ij} - q^{ij}K)  ,
\label{e0a}
\end{align}
where $K_{ij}$ is the extrinsic curvature of a slice of constant $t$.
I denote the spatial covariant derivative compatible with $q_{ij}$ 
by $D_i$, and lower and raise indices with $q_{ij}$ and its inverse
$q^{ij}$.

The constraints are generated by the Hamiltonian
\begin{align}
H[\xi^\perp,{\hat\xi}^i] = \int_\Sigma d^{n-1}x \left(
         \xi^\perp\mathcal{H} + {\hat\xi}^i\mathcal{H}_i\right)  ,
\label{e1}
\end{align}
where 
\begin{align}
\mathcal{H} = \frac{1}{\sqrt{q}}\left(\pi^{ij}\pi_{ij} - \frac{1}{n-2}\pi^2\right)
     - \sqrt{q}\,{}^{(n-1)}\!R, \qquad
\mathcal{H}^i = -2D_j\pi^{ij}  .
\label{e2}
\end{align}
Here, ${}^{(n-1)}\!R$ is the spatial curvature scalar, and the integral is over 
a Cauchy surface $\Sigma$.\footnote{I use the sign conventions of \cite{Waldb}, 
and units $16\pi G=1$, although I will occasionally restore factors of $G$.}  
The transformation generated by (\ref{e1}) is not a diffeomorphism, but is 
equivalent on shell to the diffeomorphism generated by the vector field 
$\xi^\mu$ given in (\ref{b4}).

On a manifold without a spatial boundary, the generators have Poisson brackets
\begin{align}
\left\{ H[\xi], H[\eta]\right\} = H[\{\xi,\eta\}_{\scriptscriptstyle\mathit{SD}}]
\label{e4}
\end{align}
where the surface deformation brackets 
$\{\cdot,\cdot\}_{\scriptscriptstyle\mathit{SD}}$ are given by (\ref{b5}).
As noted above, for a manifold with boundary one must add boundary 
terms to make the variation of the constraints well-defined.  The resulting
modification of the Poisson algebra (\ref{e4}) depends on the exact choice 
of boundary conditions.  

But even without these details, one can learn a good deal.
The key observation is that, by construction, the new generator 
${\bar H}[\xi] = H[\xi] + B[\xi]$ has a variation with no boundary 
contributions.  Their Poisson bracket is thus simply the standard one,
\begin{align}
\left\{ {\bar H}[\xi],{\bar H}[\eta]\right\} = \int d^{n-1}x\left(
     \frac{\delta{\bar H}[\xi]}{\delta q_{ij}}
                            \frac{\delta{\bar H}[\eta]}{\delta \pi^{ij}}
   -\frac{\delta{\bar H}[\eta]}{\delta q_{ij}}
                            \frac{\delta{\bar H}[\xi]}{\delta \pi^{ij}} \right)  .
\label{e5}
\end{align}
From (\ref{b7}), the central term is thus
\begin{align}
K[\xi,\eta] &= \left\{ {\bar H}[\xi], {\bar H}[\eta] \right\} 
    - {\bar H}[\{\xi,\eta\}_{\scriptscriptstyle\mathit{SD}}] \nonumber\\
     &= \int d^{n-1}x\left(
     \frac{\delta{\bar H}[\xi]}{\delta q_{ij}}
                            \frac{\delta{\bar H}[\eta]}{\delta \pi^{ij}}
   -\frac{\delta{\bar H}[\eta]}{\delta q_{ij}}
                            \frac{\delta{\bar H}[\xi]}{\delta \pi^{ij}} \right) 
    - H[\{\xi,\eta\}_{\scriptscriptstyle\mathit{SD}}] 
    - B[\{\xi,\eta\}_{\scriptscriptstyle\mathit{SD}}]  .
\label{e6}
\end{align}

The variational derivatives in (\ref{e6}) can be rearranged to give 
$H[\{\xi,\eta\}_{\scriptscriptstyle\mathit{SD}}]$, but only after integration
by parts.  A tedious but straightforward calculation yields
\begin{alignat}{3}
K[\xi,\eta] = &-B[\{\xi,\eta\}_{\scriptscriptstyle\mathit{SD}}] \label{e7}\\
    &-\frac{1}{8\pi G}\int_{\partial\Sigma}d^{n-2}x \sqrt{\sigma}\,n^k \Bigl[
    &&-\frac{1}{2}\frac{1}{\sqrt{q}}
               ({\hat\xi}_k\eta^\perp - {\hat\eta}_k\xi^\perp)\mathcal{H}
     - ({\hat\xi}_i\eta^\perp - {\hat\eta}_i\xi^\perp)\,{}^{(n-1)}\!R^i{}_k
    \nonumber\\
    &&& + (D_i{\hat\xi}_kD^i\eta^\perp - D_i{\hat\eta}_kD^i\xi^\perp)
    - (D_i{\hat\xi}^iD_k\eta^\perp - D_i{\hat\eta}^iD_k\xi^\perp) \nonumber\\
    &&& + \frac{1}{\sqrt{q}}
     \left({\hat\eta}_k\pi^{mn}D_m{\hat\xi}_n
              - {\hat\xi}_k\pi^{mn}D_m{\hat\eta}_n\right)
     +\frac{1}{\sqrt{q}}\pi_{ik}\{\xi,\eta\}_{\scriptscriptstyle\mathit{SD}}^i
       \Bigr]  , \nonumber  
\end{alignat}
where $\sigma_{ij}$ is the induced metric on $\partial\Sigma$ and $n^k$ is
the unit normal to $\partial\Sigma$.  

The central term (\ref{e7}) depends 
on the boundary contribution $B[\{\xi,\eta\}_{\scriptscriptstyle\mathit{SD}}]$.  
But for the computation of the central charge of a Virasoro subalgebra, 
we do not need to know this term (although we ought to check that it is finite).  
Indeed, the boundary term depends on the deformation parameters 
only in the combination $\{\xi,\eta\} = \xi\eta' - \eta\xi'$.  The central term
in a Virasoro algebra (\ref{c2}), in contrast, depends on the combination 
$\xi'\eta'' - \eta'\xi''$, which cannot be built from $\{\xi,\eta\}$ and its 
derivatives.  If such a structure appears in (\ref{e7}), it is thus necessarily 
a genuine central term.

To compute black hole entropy from the canonical version (\ref{d2}) of
the Cardy formula, it is sufficient to know the central charge and the 
temperature.  To use the microcanonical form (\ref{d1}), on the other
hand, one needs the conformal weight, which is obtained from the boundary
term $B[\xi]$.  This term does not have a simple, universal form---it depends
on the choice of boundary conditions---but it must be chosen to cancel the
boundary variation of $H[\xi]$.  This variation is \cite{Brownb}
\begin{align}
\delta H[\xi] = \dots - \frac{1}{16\pi G}
    \int_{\partial\Sigma}d^{n-2}x\Bigl\{\sqrt{\sigma}\Bigl[&
    \xi^\perp(n^k\sigma^{\ell m} - n^m\sigma^{\ell k})D_m\delta q_{k\ell}
    \label{e8}\\
    &- D_m\xi^\perp(n^k\sigma^{\ell m} - n^m\sigma^{\ell k})\delta q_{k\ell}
     \Bigr] 
     + 2{\hat\xi}^i\delta\pi^n{}_i - {\hat\xi}^n\pi^{ij}\delta q_{ij} \Bigr\}  .
     \nonumber  
\end{align}
Note that this variation determines the boundary term only up to an 
additive constant, which must be fixed by the physics and can 
sometimes be troublesome.

\section{Some examples \label{seceg}}

With these general results in hand, we can now look at a few concrete
examples.

\subsection{The BTZ black hole}

The BTZ metric \cite{BTZ} describes a (2+1)-dimensional asymptotically anti-de
Sitter black hole.  It is a stationary, axisymmetric solution of the vacuum 
Einstein field equations with a cosmological constant $\Lambda = -1/\ell^2$, 
and is given in ADM form by
\begin{align}
&ds^2 = N^2dt^2 - N^{-2}dr^2
  - r^2\left( d\varphi + N^\varphi dt\right)^2
\label{e9}
\intertext{with}
&N = \left( -8GM + \frac{r^2}{\ell^2} + \frac{16G^2J^2}{r^2} \right)^{1/2} ,
  \quad N^\varphi = - \frac{4GJ}{r^2} \qquad  (|J|\le M\ell)  .
\label{e10}
\end{align}
The only nonvanishing component of the ADM momentum is
\begin{align}
\pi^{r\varphi} = \frac{r_+r_-}{\ell r^2}  .
\label{e11}
\end{align}
Pure anti-de Sitter space is described by a similar metric, with $J=0$ and
$8GM=-1$.
 
Like every solution of the vacuum field equations in 2+1 dimensions, the
BTZ black hole has constant curvature.  Nevertheless, somewhat surprisingly, 
it is a true black hole.  In particular, it has an event horizon at $r_+$ and, when 
$J\ne0$, an inner Cauchy horizon at $r_-$, where
\begin{align}
r_\pm^2=4GM\ell^2\left ( 1 \pm
\left [ 1 - \left(\frac{J}{M\ell}\right )^2\right ]^{1/2}\right ) , \qquad
\hbox{i.e.,}\qquad
M=\frac{r_+^2+r_-^2}{8G\ell^2}, \quad J=\frac{r_+ r_-}{4G\ell}  .
\label{e12}
\end{align}
The horizon rotates at an angular velocity
\begin{align}
\Omega_H = -N^\varphi(r_+) = \frac{r_-}{r_+\ell}  .
\label{e12a}
\end{align}
Kruskal coordinates were derived in \cite{BHTZ}; the Penrose diagram
is essentially the same as that of an asymptotically anti-de Sitter black hole 
in 3+1 dimensions.  

In 1986, Brown and Henneaux stumbled across an aspect of what we would 
now call the AdS/CFT correspondence \cite{BrownH}.\footnote{The 
full AdS/CFT correspondence involves many additional degrees of freedom;
it remains unclear whether (2+1)-dimensional gravity alone contains 
enough degrees of freedom to fully account for black hole entropy
\cite{Carlipb,Witten,Maloney}.}  Analyzing 
the asymptotic symmetries of general relativity in (2+1)-dimensional 
asymptotically anti-de Sitter space, they found a pair of commuting Virasoro 
algebras---the symmetry group of a two-dimensional conformal  field theory%
---with central charges
\begin{align}
c^\pm = \frac{3\ell}{2G}  .
\label{e13}
\end{align}
To obtain this result, Brown and Henneaux imposed boundary conditions
\begin{align}
g_{ab} \sim \left( \begin{array}{ccc}
     -\frac{r^2}{\ell^2} + \mathcal{O}(1) & \mathcal{O}(\frac{1}{r^3}) 
                          & \mathcal{O}(1)\\[1ex]
     & \frac{\ell^2}{r^2} + \mathcal{O}(\frac{1}{r^4}) 
                          & \mathcal{O}(\frac{1}{r^3})\\[1ex]
     && r^2 + \mathcal{O}(1) \end{array}\right)  
\label{e14}
\end{align}
at spatial infinity.  (My index order is $(t,r,\varphi)$.)  These conditions are 
satisfied by the BTZ black hole, as well as by pure anti-de Sitter space and 
anti-de Sitter space with point particles.  It is straightforward to check that the 
diffeomorphisms that respect these boundary conditions are generated by 
two families of vector fields $\xi^\pm$, with
\begin{align}
\xi^{\pm t} &= \frac{\ell}{2}\left(T^\pm 
    +\frac{\ell^2}{2r^2}\partial_\varphi^2T^\pm\right) \nonumber\\
\xi^{\pm r} &= \mp\frac{r}{2}\partial_\varphi T^\pm \label{e15}\\
\xi^{\pm \varphi} &= \pm\frac{1}{2}\left(T^\pm 
    -\frac{\ell^2}{2r^2}\partial_\varphi^2T^\pm\right)   , \nonumber
\end{align}
where $T^\pm = T^\pm(\varphi\pm t/\ell)$ are left- and right-moving modes 
on the asymptotic boundary.  

The vector fields (\ref{e15}) are normalized so that
\begin{align}
&\{ \xi^\pm[T_1^\pm], \xi^\pm[T_2^\pm] \} 
    = \xi^\pm[T_1^\pm\partial_\varphi T_2^\pm - T_2^\pm\partial_\varphi T_1^\pm]
   \nonumber\\
&\{\xi^+[T_1^+], \xi^-[T_2^-] \} = 0
\label{e16}
\end{align}
From (\ref{c1})--(\ref{c2}), the corresponding generators of surface deformations 
should form a pair of commuting Virasoro algebras.  To find the central charges,
we can use the general result (\ref{e7}).  The relevant three-derivative terms are
\begin{align}
K[\xi,\eta] = \dots
    -\frac{1}{8\pi G}\int_{\partial\Sigma}\!\!d\varphi\ rN& \Bigl[
    - ({\hat\xi}_r\eta^\perp - {\hat\eta}_r\xi^\perp)\,{}^{(2)}\!R^r{}_r \label{e17}\\
    &+ (D_\varphi{\hat\xi}_rD^\varphi\eta^\perp - D_\varphi{\hat\eta}_rD^\varphi\xi^\perp)
    - (D_\varphi{\hat\xi}^\varphi D_r\eta^\perp - D_\varphi{\hat\eta}^\varphi D_r\xi^\perp)  
       \Bigr]  , \nonumber
\end{align}
where $D_\varphi{\hat\xi}^r = \partial_\varphi{\hat\xi}^r - rN^2{\hat\xi}^\varphi$
and $D_\varphi{\hat\xi}^\varphi = \partial_\varphi{\hat\xi}^\varphi + \frac{1}{r}{\hat\xi}^r$
are spatial covariant derivatives.  A straightforward calculation yields
\begin{align}
K[\xi^\pm[T^\pm_1],\xi^\pm[T^\pm_2]] = \dots + \frac{\ell}{32\pi G} 
   \int_{\partial\Sigma}\!\!d\varphi 
   \left( \partial_\varphi T^\pm_1\partial_\varphi^2T^\pm_2
      - \partial_\varphi T^\pm_2\partial_\varphi^2T^\pm_1\right)  .
\label{e18}
\end{align}
Comparing this to the general Virasoro algebra (\ref{c2}), we can read off 
the central charges, confirming the results (\ref{e13}).

To use the microcanonical Cardy formula, we also need the boundary term $B[\xi]$.
This was also computed by Brown and Henneaux.  From the boundary conditions
(\ref{e14}), it is easy to see that the only nonvanishing terms in the variation 
(\ref{e9}) are
\begin{align}
\delta H[\xi] &= \dots - \frac{1}{16\pi G}
    \int d\varphi\left\{rn^r\sigma^{\varphi\varphi}\left[ 
    \xi^\perp( D_\varphi\delta q_{r\varphi} - D_r\delta q_{\varphi\varphi}\right]
     + rn^r\sigma^{\varphi\varphi}D_r\xi^\perp \delta q_{\varphi\varphi}
     + 2{\hat\xi}^\varphi\delta\pi^r{}_\varphi\right\} \nonumber\\
    &= \dots -  \frac{1}{16\pi G} \int d\varphi \left[
    \xi^t\delta\left(\frac{r_+^2+r_-^2}{\ell^2}\right) 
    + {\hat\xi}^\varphi\delta\left(\frac{2r_+r_-}{\ell}\right)\right]
    = \dots -\frac{1}{2\pi}\delta\int d\varphi \left[
    \xi^t M + \xi^\varphi  J\right]  ,
\label{e19}
\end{align}
justifying the notation $M$ and $J$ for the mass and angular momentum 
of the black hole.  Evaluated on the asymptotic symmetries (\ref{e15}), the 
boundary term is then
\begin{align}
B[T^\pm] = \frac{(r_+\pm r_-)^2}{32\pi G\ell}\int d\varphi\, T^\pm + B_0  .
\label{e20}
\end{align}
In particular, for the zero modes $T_0^\pm=1$, we have
\begin{align}
\Delta^\pm = B[T_0^\pm] = \frac{(r_+\pm r_-)^2}{16 G\ell} + \frac{\ell}{16G}  ,
\label{e21}
\end{align}
where the constant $B_0$ has been fixed by demanding that anti-de Sitter space
have $\Delta^\pm=0$.  As Strominger \cite{Strominger} and Birmingham, 
Sachs, and Sen \cite{BSS} observed, these conformal weights and the central 
charges (\ref{e13}) then determine an entropy
\begin{align}
S = 2\pi\sqrt{\frac{c^+}{6}\left(\Delta^+ - \frac{c^+}{24}\right)} 
   + 2\pi\sqrt{\frac{c^-}{6}\left(\Delta^- - \frac{c^-}{24}\right)}
   = \frac{2\pi r_+}{4G}  ,
\label{e22}
\end{align}
one quarter of the horizon ``area'' $2\pi r_+$, matching the standard 
Bekenstein-Hawking expression.

It is interesting to consider the canonical version (\ref{d2}) of the Cardy 
formula as well.  The BTZ black hole has a Hawking temperature
\begin{align}
T_H = \frac{\kappa}{2\pi} = \frac{r_+^2-r_-^2}{2\pi\ell r_+}  .
\label{e23}
\end{align}
The inverse of this temperature gives the periodicity of Greens functions
in imaginary time.  For our two Virasoro algebras, though, the relevant
periodicities are in the coordinates $t^\pm = t\pm\ell\varphi$ that
appear in the asymptotic diffeomorphisms.  To obtain the corresponding
``left'' and ``right'' temperatures, following \cite{Guica}, we first note 
that the modes of a scalar field in a BTZ background take the form
\begin{align}
\phi \sim F(r) e^{im\varphi-i\omega t} 
     = F(r)e^{-i\omega^+t^+-i\omega^-t^-} \quad
\hbox{with}\quad
\omega^\pm = \frac{1}{2}\left(\omega \mp \frac{m}{\ell}\right)  .
\label{e24}
\end{align}
The standard Boltzmann factor depends on both the energy and the angular
velocity $\Omega_H$, and can be written as
\begin{align}
e^{-\beta_H(\omega-m\Omega_H)} = e^{-\beta^+\omega^+ - \beta^-\omega^-}
\quad\hbox{with}\quad
\beta^\pm = \beta(1\pm\Omega_H\ell) = \frac{2\pi\ell}{r_+\mp r_-}  .
\label{e26}
\end{align}
Substituting the corresponding temperatures $T^\pm = 1/\beta^\pm$ into 
the canonical Cardy formula, we obtain
\begin{align}
S = \frac{\pi^2}{3}c^+T^+ + \frac{\pi^2}{3}c^-T^-
   = \frac{2\pi r_+}{4G}  ,
\label{e27}
\end{align}
again giving the standard Bekenstein-Hawking entropy.

In some ways, the BTZ black hole is a very special case: the asymptotic
boundary is a two-dimensional cylinder, so an asymptotic conformal field
theory seems quite natural.  But the result also has a simple extension.
Many near-extremal black holes have a near-horizon geometry of the form 
$\mathit{BTZ}\times\mathit{trivial}$, and the results of this section
can be applied almost unchanged, except for appropriate redefinitions of
constants.  For a good review, see \cite{Skenderis}.

\subsection{The extremal Kerr black hole \label{secextremal}}

As our next example, let us consider the near-horizon geometry of the
extremal Kerr black hole \cite{Guica,Bredbergb}, a system that has
attracted a good deal of recent attention.  In Boyer-Lindquist 
coordinates, the ADM form of the extremal Kerr metric is \cite{Frolov}
\begin{align}
ds^2  = -N^2dt^2  + \frac{\Sigma}{\Delta}dr^2
      + \frac{A\sin^2\theta}{\Sigma}\left(d\varphi + N^\varphi dt\right)^2 
      + \Sigma\, d\theta^2   ,
\label{f1}
\end{align}
where 
\begin{align}
&\Delta = (r-r_+)^2, \qquad
\Sigma = r^2 + r_+{}^2\cos^2\theta, \qquad
A = (r^2 + r_+{}^2)^2 - (r-r_+)^2r_+{}^2\sin^2\theta 
\nonumber\\
&N = \sqrt{\frac{\Sigma\Delta}{A}} ,
 \qquad N^\varphi = \frac{2r_+{}^2r}{A}  .
\label{f2}
\end{align}
This black hole has a mass $GM=r_+$, an angular momentum $J=r_+^2/G$,
and a horizon at $r=r_+$ that rotates with an angular velocity
\begin{align}
\Omega_H = N^\varphi(r_+) = \frac{1}{2r_+}
\label{f3}
\end{align}

At extremality, the Kerr black hole has an infinite throat: that is, the proper 
distance from the horizon $r_+$ to any point $r>r_+$ is infinite.  To focus
in on the near-horizon region, Bardeen and Horowitz considered a coordinate
transformation \cite{Bardeen}
\begin{align}
{\bar t} = \frac{\lambda t}{2m}, \quad y = \frac{\lambda m}{r-m}, 
\quad {\bar\varphi} = \varphi - \Omega_Ht
\label{f5}
\end{align}
followed by a limit $\lambda\rightarrow0$.   The resulting near-horizon 
extremal Kerr, or NHEK, metric is
\begin{align}
ds^2 = r_+^2(1+\cos^2\theta)\left( \frac{-d{\bar t}^2 + dy^2}{y^2} 
     + \Lambda^2\left(d{\bar\varphi} + \frac{d{\bar t}}{y}\right)^2 + d\theta^2\right)
\label{f6}
\end{align}
with
\begin{align}
\Lambda = \frac{2\sin\theta}{1+\cos^2\theta}  .
\label{f7}
\end{align}
The only nonvanishing component of the canonical momentum is
\begin{align}
\pi^{y{\bar\varphi}} = \frac{\sqrt{q}}{2N}q^{yy}\partial_yN^{\bar\varphi}
   = -\frac{\Lambda}{2}  .
\label{f4}
\end{align}

Guica et al.\ consider boundary conditions \cite{Guica}
\begin{align}
\delta g_{ab} \sim \left(\begin{array}{cccc}
   \mathcal{O}(\frac{1}{y^2}) & \mathcal{O}(1) & \mathcal{O}(y) & \mathcal{O}(1)\\
   & \mathcal{O}(\frac{1}{y}) &\mathcal{O}(1) & \mathcal{O}(\frac{1}{y})\\
   &&\mathcal{O}(y) &\mathcal{O}(y)\\
   &&&\mathcal{O}(1)
   \end{array}\right)
\label{f7a}
\end{align}
near the asymptotic boundary $y=0$. (My index order is 
$({\bar t},y,\theta,{\bar\varphi})$.) These conditions, which are strong enough
to lead to finite charges, are preserved by the 
diffeomorphisms
\begin{align}
\xi^{\bar\varphi} = \epsilon({\bar\varphi}) , \qquad
\xi^y = y\partial_{\bar\varphi}\epsilon({\bar\varphi})  ,
\label{f8}
\end{align}
which satisfy a $\operatorname{Diff}S^1$ algebra.  (Constant time translations
are also permitted,  but are not relevant to the dual conformal field theory.)
Since (\ref{f8}) includes no $\xi^\perp$ component, the sole contributions 
to the central term (\ref{e7}) come from the terms involving the canonical 
momentum $\pi^{ij}$.  The only three-derivative term is
\begin{align} 
K[\xi,\eta] &= \dots -\frac{1}{8\pi G}\int d^2x  
     \left({\hat\eta}^y\pi^{mn}D_m{\hat\xi}_n
              - {\hat\xi}^y\pi^{mn}D_m{\hat\eta}_n\right) \nonumber\\
     &= \dots + \frac{1}{8\pi G}\int d\theta d{\bar\varphi}\,
     \frac{\Lambda}{2}\left(\eta^yq_{yy}\partial_{\bar\varphi}\xi^y
     - \xi^yq_{yy}\partial_{\bar\varphi}\eta^y\right)  \nonumber\\
     &= \dots - \frac{r_+^2}{16\pi G}\int d\theta \Lambda(1+\cos^2\theta)
     \int d{\bar\varphi}\left(
     \partial_{\bar\varphi}\eta^{\bar\varphi}\partial_{\bar\varphi}^2\xi^{\bar\varphi}
     -\partial_{\bar\varphi}\xi^{\bar\varphi}\partial_{\bar\varphi}^2\eta^{\bar\varphi}
     \right)\nonumber\\ 
     &= \dots + \frac{r_+^2}{4\pi G}\int d{\bar\varphi}\left(
     \partial_{\bar\varphi}\xi^{\bar\varphi}\partial_{\bar\varphi}^2\eta^{\bar\varphi}
     -\partial_{\bar\varphi}\eta^{\bar\varphi}\partial_{\bar\varphi}^2\xi^{\bar\varphi}
     \right)  .
\label{f9}
\end{align}
Comparing to (\ref{c2}), we see that we one again have a Virasoro algebra,  
with central charge
\begin{align}
c = \frac{12r_+^2}{G} = 12J  .
\label{f10}
\end{align}

There is an important difference between this central charge and that of the BTZ
black hole.  For the BTZ black hole, $c$ depends only on the cosmological 
constant, and not on the particular parameters of the black hole.  It is
thus plausible to conjecture that a single conformal field theory encompasses
all (2+1)-dimensional black holes.  Here, in contrast, $c$ depends on $J$, 
implying a different dual conformal field theory for each value of angular 
momentum.  

This may not be important for extremal rotating black holes, which have 
a vanishing temperature and do not undergo Hawking radiation.  But  
let us suppose---as suggested, for instance, in \cite{Castrob}---that a 
single conformal field theory also describes nonextremal black holes with 
the same angular momentum.  Then Hawking radiation, which will typically 
change the value of $J$, will take us from one conformal dual to another.
This suggests that a full treatment of Hawking radiation, including its
back-reaction on the black hole, may involve flows between conformal field
theories.

To use the microcanonical Cardy formula, we need the boundary term $B[\xi]$.  
Again, the absence of a $\xi^\perp$ component in the deformation (\ref{f8}) 
makes this easy to compute: the only relevant term in (\ref{e8}) is
\begin{align}
\delta H[\xi] &= \dots - \frac{1}{16\pi G}
    \int_{\partial\Sigma}d\theta d{\bar\varphi}
   \left( 2{\hat\xi}^{\bar\varphi}\delta\pi^y{}_{\bar\varphi}\right) \nonumber\\
   &= \dots - \frac{1}{16\pi G} \int d{\bar\varphi}\,\epsilon({\bar\varphi})
    \int d\theta \left(-\frac{\Lambda}{2}\cdot r_+^2(1+\cos^2\theta)\Lambda^2\right)
   \nonumber\\
   &= \dots + \frac{r_+^2}{2\pi G} \int d{\bar\varphi}\,\epsilon({\bar\varphi})
   \int d\theta \frac{\sin^3\theta}{(1+\cos^2\theta)^2}
   = \dots + \frac{r_+^2}{G} \int \frac{d{\bar\varphi}}{2\pi}\,\epsilon({\bar\varphi})  .
\label{f11}
\end{align}
If we now take the undetermined constant $B_0$ in the boundary term to be
zero---a plausible assumption, but not one that is obviously true---the Cardy 
formula yields
\begin{align}
S = 2\pi\sqrt{\frac{c}{6}\left(\Delta - \frac{c}{24}\right)} = \frac{2\pi r_+^2}{G}  ,
\label{f12}
\end{align}
which is the correct Bekenstein-Hawking entropy for the extremal Kerr black hole.

We can also use the canonical version of the Cardy formula.  This is a bit subtle: the
Hawking temperature for a general Kerr black hole is
\begin{align} 
T = \frac{r_+-M}{4\pi Mr_+}  ,
\label{f13}
\end{align}
which vanishes for an extremal black hole.  Following Guica et al.\ \cite{Guica}, 
though, we can obtain an appropriate temperature as a limit.  For a
\emph{nonextremal} black hole, the modes of the Frolov-Thorne vacuum
\cite{Frolovb} are of the form
\begin{align}
\phi \sim F(r,\theta)e^{im\varphi-i\omega t} \sim
   {\bar F}(y,\theta)e^{in_L{\bar\varphi}-in_R{\bar t}}\quad
\hbox{with}\quad
n_L = m, \ n_R = \frac{1}{\lambda}(2M\omega - m)  ,
\label{f14}
\end{align}
where I have performed the coordinate transformation (\ref{f5}) but not
yet taken the $\lambda\rightarrow0$ limit.  As in the BTZ case, we can
compute the corresponding Boltzmann factor,
\begin{align}
e^{-\beta_H(\omega - m\Omega_H)} = e^{-\beta_R n_R - \beta_Ln_L}
\quad\hbox{with}\quad
\beta_R = 2\pi\left(\frac{r_+ - 2\Omega_Hr_+M}{r_+-M}\right), \quad 
\beta_L = \frac{2\pi\lambda r_+}{r_+-M}  .
\label{f15}
\end{align}
In the extremal limit, $r_+\rightarrow M$, the right temperature remains
finite,
\begin{align}
T_R = 1/\beta_R \rightarrow \frac{1}{2\pi}  .
\label{f16}
\end{align}
The canonical Cardy formula (\ref{d2}) then gives
\begin{align}
S = \frac{\pi^2}{3}cT = \frac{2\pi r_+^2}{G}  ,
\label{f17}
\end{align}
again matching the Bekenstein-Hawking entropy.

A similar analysis has been performed for other extremal black holes 
(see, for example, \cite{Hartman,Lu}), and has been repeated for the 
extremal Kerr solution without taking the near-horizon limit \cite{Carlipg}. 
But although some attempts have been made to extend the method 
to nonextremal black holes \cite{Castro,Rasmussen}, the situation 
remains murky.  So for the nonextremal case, it is worthwhile to look 
for an alternative approach.

\subsection{A generic (3+1)-dimensional black hole}

In the preceding examples, we looked at asymptotic boundaries far
from the black hole horizon: at infinity for the BTZ black hole, and at
the ``infinity'' of the near-horizon region for the extremal Kerr black
hole.  But as noted in section \ref{secSteps}, it is also natural to look 
at boundary conditions at the horizon itself.    

Consider an arbitrary stationary black hole in 3+1 dimensions \cite{Carlipg}.  
Such a black hole is necessarily axisymmetric \cite{Hawkingb}: that is,
it admits two Killing vectors $T^a$ and $\Phi^a$ that generate time
translations and rotations.  For the near-horizon metric, we can write 
a general ADM form
\begin{align}
ds^2 = -N^2dt^2  + d\rho^2 
      + q_{\varphi\varphi}\left(d\varphi + N^\varphi dt\right)^2 
      + q_{zz}dz^2  ,
\label{g1}
\end{align}
where $\rho$ is the proper distance from the horizon and $z$ is an
angular coordinate such as $\cos\theta$.   Even before imposing any field 
equations, it can be shown that the requirement of finite curvature at 
the horizon sharply restricts the behavior of the metric: it must have 
a near-horizon expansion \cite{Medved}
\begin{alignat}{3}
&N =  \kappa_{H}\rho 
       + \frac{1}{3!}\kappa_2(z)\rho^3 + \dots \qquad\quad
&& q_{\varphi\varphi} = [q_H]_{\varphi\varphi}(z) 
                    +  \frac{1}{2}[q_2]_{\varphi\varphi}(z)\rho^2 + \dots
       \nonumber\\[1ex]
&N^\varphi = -\Omega_{H} 
       - \frac{1}{2}\omega_2(z)\rho^2 + \dots 
&&q_{zz} = [q_H]_{zz}(z) + \frac{1}{2}[q_2]_{zz}(z)(\rho^2) + \dots ,
\label{g2}  
\end{alignat}
where the surface gravity $\kappa_{H}$ and 
the horizon angular velocity $\Omega_{H}$
are constants.  The only nonzero component of the canonical momentum
$\pi^{ij}$ near the horizon is then
\begin{align}
\pi^{\rho\varphi}  
    =  -\frac{\omega_2}{2\kappa_{H}}\sqrt{q} 
   + \mathcal{O}(\rho^2) .
\label{g3}
\end{align}

Now, for a stationary black hole---or, more generally, a black hole with a
stationary neighborhood around an isolated horizon \cite{Date}---the 
horizon $\mathcal{H}$ is a Killing horizon.  This means there exists 
a Killing vector
\begin{align}
\chi^a = T^a + \Omega_{H}\Phi^a
\label{g4}
\end{align}
that is null at $\mathcal{H}$ and is normal to $\mathcal{H}$.  Ideally, 
this is the place we would impose boundary conditions.  As noted earlier, 
though, this has so far proven very difficult.  We can, however, ``stretch'' 
the horizon into a timelike boundary $\mathcal{H}_s$ just outside the 
true horizon, determine a conformal dual there, and then take the limit 
as $\mathcal{H}_s$ approaches $\mathcal{H}$.  

In particular, let us ``stretch'' the Killing vector, by requiring that a 
new Killing vector
\begin{align} 
{\bar\chi}^a =  T^a + {\bar\Omega}\Phi^a
\label{g5}
\end{align}
be null at the intersection of the stretched horizon with our Cauchy surface
$\Sigma$.   For the metric (\ref{g1}), this is the requirement that
\begin{align}
{\bar\chi}^2 = 0 
       =-N^2 + q_{\varphi\varphi}(N^\varphi + {\bar\Omega})^2
       = -N^2 + q_{\varphi\varphi}
          (\Omega_{H} - {\bar\Omega})^2
       + \mathcal{O}(\rho^3)  .
\label{g6}
\end{align}
The parameter
\begin{align}
{\bar\varepsilon} = \Omega_{H} - {\bar\Omega}  ,
\label{g7}
\end{align}
which measures the ``stretching'' of the Killing vector ${\bar\chi}^a$,
is of order $\rho$---essentially the proper distance from the stretched horizon 
to the true horizon---as is standard in the membrane paradigm \cite{Thorne}.

At the stretched horizon $\mathcal{H}_s$, we can independently specify 
the surface deformation parameters $(\xi^\perp,{\hat\xi}^i)$ and their 
normal derivatives.   The transformations that leave the metric (\ref{g1}) 
fixed at the boundary are then  
\begin{alignat}{3}
&\delta_\xi N = 0 \quad && \Rightarrow\ \
   {\hat\xi}^\rho = -{\bar\varepsilon}\rho\partial_\varphi\xi^t 
    = -\rho{\bar\partial}_t\xi^t\nonumber\\
& \delta_\xi N^\rho = 0 && \Rightarrow\ \
    \rho\partial_\rho\xi^t = -\frac{{\bar\varepsilon}^2}
  {\kappa_{H}^2}\partial_\varphi{}^2\xi^t
   = -\frac{1}{\kappa_{H}^2}{\bar\partial}_t{}^2\xi^t
   \nonumber\\
&\delta_\xi N^\varphi = 0 && \Rightarrow\ \
   {\hat\xi}^\varphi = \frac{\kappa_{H}^2\rho^2}
   {\bar\varepsilon}q^{\varphi\varphi}\xi^t =  {\bar\varepsilon}\xi^t\nonumber\\
&\delta q_{\rho\varphi}=0 && \Rightarrow\ \
    \rho\partial_\rho{\hat\xi}^\varphi 
    = {\bar\varepsilon}\rho^2q^{\varphi\varphi}\partial_{\varphi}{}^2\xi^t 
    - \omega_2\rho^2\xi^t
    = \frac{{\bar\varepsilon}}{\kappa_{H}^2}
    {\bar\partial}_t{}^2\xi^t - \omega_2\rho^2\xi^t
\label{g8}
\end{alignat}
where $\xi^t = \xi^t(\varphi-{\bar\Omega}t)$ and
\begin{align}
{\bar\partial}_t =  \partial_t - N^i\partial_i  
     \sim \partial_t + \Omega_{H}\partial_\varphi 
\label{g9}
\end{align}
is a convective derivative.

In our earlier examples, the surface deformations were functions of an 
angular coordinate, and there was a natural choice of periodicity of the
modes.  Here, the relevant variable is really an affine parameter along 
the Killing vector, and the choice is not unique.  A particularly nice choice
of modes is
\begin{align}
\xi^t_n 
    = \frac{1}{2}\,e^{in(\varphi - {\bar\Omega}t)/{\bar\varepsilon}}  ,
\label{g10}
\end{align}
which are chosen in order that ${\bar\partial}_t{\tilde\xi}^t_n 
= in{\tilde\xi}^t_n$.   The resulting frequencies blow up at the horizon, 
but this is the ordinary blue shift as seen by an outside observer: a 
corotating observer with four-velocity $u^a = (T^a - N^\varphi\Phi^a)/N$ 
will see a frequency $k_au^a = n/N(1+\mathcal{O}(\rho))$, the  
usual blue shift factor.  Other choices of moding are discussed in 
\cite{Carlipg}; they typically lead to different, and sometimes divergent,
central charges and conformal weights, but to the same entropy.

The normalization of (\ref{g10}) has been chosen to ensure that the 
surface deformation brackets (\ref{b5}) of these modes are consistent, 
with\footnote{Here and in what follows, there is a subtlety regarding
the radial derivatives $\partial_\rho$, coming from the fact that the
proper distance $\rho$ is metric-dependent.  This issue is discussed in 
detail in the appendices of \cite{Carlipg}.}
\begin{align}
\{{\xi}_m,{\xi}_n\}_{\scriptscriptstyle\mathit{SD}}^t =  i(m-n)\xi_{m+n}^t  .
\label{g11}
\end{align}
We thus expect a Virasoro algebra to once again be present.  The relevant 
contributions to the central term (\ref{e7}) are now
\begin{align}
K[\xi,\eta] &= -\frac{1}{8\pi G}\int_{\partial\Sigma}d^2x\sqrt{\sigma}\, 
   n^k \Bigl[(D_i{\hat\xi}_kD^i\eta^\perp  - D_i{\hat\xi}^iD_k\eta^\perp)
   - (\xi\leftrightarrow\eta)\Bigr] \nonumber\\
   &= -\frac{1}{4\pi G\kappa_{H}}
   \frac{\mathcal{A}}{2\pi}\int d\varphi\, 
   \left[ {\bar\partial}_t\xi^t{\bar\partial}_t{}^2\eta^t 
            - {\bar\partial}_t\eta^t{\bar\partial}_t{}^2\xi^t\right]  ,
\label{g12}
\end{align}
where $\mathcal{A}$ is the horizon area.  Here I have used (\ref{g8}) for 
both the components of $\xi$ and $\eta$ and their radial derivatives, along
with the identity
\begin{align}
\int d^2x\,\sqrt{\sigma}F(\varphi) 
   = \frac{\mathcal{A}}{2\pi}\int d\varphi\, F(\varphi)  ,
\label{g13}
\end{align}
which follows from the fact that the boundary metric $\sigma_{ij}$ is
independent of $\varphi$.  The expression (\ref{g12}) may again be 
recognized as the central term of a Virasoro algebra, with central charge
\begin{align}
c = \frac{3\mathcal{A}}{2\pi G\kappa_{H}}  .
\label{g14}
\end{align}

It is now easy to use the canonical form of the Cardy formula.  The Hawking
temperature of a black hole with metric (\ref{g1}) is
\begin{align}
T_ {H}
   = \frac{\kappa_{H}}{2\pi}  .
\label{g15}
\end{align}
The coordinate along our stretched horizon is 
$(\varphi - {\bar\Omega}t)/{\bar\epsilon}$, so the relevant modes are
\begin{align}
\phi \sim F(r,\theta)e^{im\varphi-i\omega t} \sim
    F(r,\theta)e^{in_L((\varphi-{\bar\Omega}t)/{\bar\epsilon})-in_R t}\quad
\hbox{with}\quad
n_L = m{\bar\epsilon}, \ n_R = \omega - m{\bar\Omega}  .
\label{g16}
\end{align}
The Boltzmann factor is then
\begin{align}
e^{-\beta_H(\omega - m\Omega_{H})} 
   = e^{-\beta_R n_R - \beta_Ln_L}
\quad\hbox{with}\quad
\beta_R =  \beta_{H}, \quad 
\beta_L =  
   \frac{{\bar\Omega}-\Omega_{H}}{\bar\epsilon}\beta_H
   = \beta_{H}  ,
\label{g17}
\end{align}
and the entropy is thus
\begin{align}
S = \frac{\pi^2}{3}cT_{H}
   = \frac{\mathcal{A}}{4G}  ,
\label{g18}
\end{align}
as expected.  The microcanonical version of the derivation is discussed 
in \cite{Carlipg}; it appears to also yield the correct Bekenstein-Hawking 
entropy, although the zero of the conformal weight $\Delta$ is not 
entirely clear.

While this particular result is fairly new, it is closely related to the 
covariant phase space analysis of \cite{Carlipd}.  That approach
has been applied to a wide variety of black holes, including dilaton
black holes \cite{Jing} and black holes in higher derivative theories 
\cite{Cvitan,Cvitanb}.  As in the extremal Kerr case, the central charge
depends on parameters of the black hole, and once again, black holes
with different horizon areas must be described by different conformal
field theories.

\subsection{Two-dimensional dilaton gravity}

I will conclude this section with an interesting example of a model for
which these conformal methods fail, but in an interesting way.  In
two spacetime dimensions, the Einstein-Hilbert Lagrangian is a
total derivative, and to obtain an interesting theory of gravity, one 
must add new fields.  The simplest choice is the Jackiw-Teitelboim
model \cite{Jackiw,Teitelboimb}, with an action
\begin{align}
I = \frac{1}{2}\int d^2x\,\sqrt{-g}\,\phi \left(R + \frac{2}{\ell^2}\right)  ,
\label{h1}
\end{align}
whose fields include both the metric and a dilaton $\phi$.  The
field equations obtained from this action have an asymptotically 
anti-de Sitter black hole solution,
\begin{align}
ds^2 = -\left(\frac{x^2}{\ell^2}-a^2\right)dt^2 
   + \left(\frac{x^2}{\ell^2}-a^2\right)dx^2, \quad
\phi = \frac{\phi_0 x}{\ell}  ,
\label{h2}
\end{align}
with a mass
\begin{align}
M = \frac{\phi_0a^2}{2\ell}
\label{h3}
\end{align}
and an entropy \cite{Cadoni}
\begin{align}
S = \frac{2\pi\phi_0a}{\ell}  .
\label{h4}
\end{align}

The asymptotic symmetries of this model were first analyzed by Hotta
\cite{Hotta} and Cadoni and Mignemi \cite{Cadonia,Cadonib}.  The
ADM constraints are
\begin{align}
H[\xi^\perp,\xi^x] = \int_\Sigma dx\left\{ \xi^\perp \left[
   -\pi_\phi\pi_\sigma + \left(\frac{\phi'}{\sigma}\right)' 
   - \lambda^2\sigma\phi\right] 
   + \xi^x\left[\pi_\phi \phi' - \pi_\sigma'\sigma\right]\right\}
\label{h5}
\end{align}
where $\sigma=\sqrt{q_{xx}}$.  In the language of section \ref{secADM}
of this paper, one finds that
\begin{align}
\left\{{\bar H}[\xi],{\bar H}[\eta]\right\}
  = {\bar H}[\{\xi,\eta\}_{\scriptscriptstyle\mathit{SD}}] + K(\xi,\eta)
\label{h6} 
\end{align}
with
\begin{align}
   \{\xi,\eta\}_{\scriptscriptstyle\mathit{SD}}^\perp =
          \xi^x\eta^{\perp\prime} -  \eta^x\xi^{\perp\prime}, \quad
   \{\xi,\eta\}_{\scriptscriptstyle\mathit{SD}}^x =
          \xi^x\eta^{x\prime} - \eta^x\xi^{x\prime}
         + \frac{1}{\sigma^2}\left(
         \xi^\perp\eta^{\perp\prime} - \eta^\perp\xi^{\perp\prime}\right)
\label{h7}
\end{align}
and
\begin{align}
K(\xi,\eta) = &-B[\{\xi,\eta\}_{\scriptscriptstyle\mathit{SD}}] \label{h8}\\
   &+ \left[ (\xi^x\eta^\perp - \eta^x\xi^\perp)\left(\pi_\phi\pi_\sigma 
       +\frac{\sigma\phi}{\ell^2}\right)
   - (\xi^x\eta^{\perp\prime} - \eta^x\xi^{\perp\prime})\frac{\phi'}{\sigma}
   + \xi^\perp\eta^{\perp\prime} - \eta^\perp\xi^{\perp\prime}%
        \frac{\pi_\sigma}{\sigma} \right]_{\partial\Sigma}  .  \nonumber
\end{align}
Note that the Cauchy surface $\Sigma$ is one-dimensional, so its boundary 
is a discrete set of points; there is thus no integral in (\ref{h8}).

The asymptotic conditions
\begin{align}
g_{ab} \sim \left(\begin{array}{cc} 
      -\frac{x^2}{\ell^2} + \mathcal{O}(1) &  \mathcal{O}(\frac{1}{x^3})\\[1ex]
      & \frac{\ell^2}{x^2} + \mathcal{O}(\frac{1}{x^4})
      \end{array}\right) , \quad \phi \sim \mathcal{O}(x)
\label{h9}
\end{align} 
are preserved by the diffeomorphisms 
\begin{align}
\xi^t = T + \frac{\ell^4}{2}{\ddot T}  , \quad
\xi^x = -x{\dot T}
\label{h10}
\end{align}
for arbitrary functions $T(t)$.  The central term (\ref{h8}) then has a piece of 
the form
\begin{align}
K(\xi_1,\xi_2) = \dots 
   -\ell\phi_0 \left({\dot T}_1{\ddot T}_2 - {\dot T}_2{\ddot T}_1\right) .
\label{h11}
\end{align}
This is \emph{almost} the right form for the central extension of a Virasoro
algebra.  Indeed, as Cadoni and Mignemi pointed out, if one were to arbitrarily
introduce an integral over $t$ with period $2\pi\ell$, the period of
Euclidean AdS space, one would find a central charge
\begin{align}
c = 24\phi_0  .
\label{h12}
\end{align}
This is too large by a factor of two: for instance, the temperature of the black
hole (\ref{h2}) is
\begin{align}
T_H = \frac{a}{2\pi\ell}  ,
\label{h13}
\end{align}
and the canonical Cardy formula would give
\begin{align}
S = \frac{\pi^2}{3}cT = \frac{4\pi\phi_0a}{\ell}  ,
\label{h14}
\end{align}
twice the entropy (\ref{h4}).  One can also compare (\ref{h12}) with
the central charge (\ref{g14}) of the preceding section, noting that the
dimensional reduction of (3+1)-dimensional spherically symmetric gravity 
gives dilaton gravity with a dilaton $\phi =  {\mathcal{A}}/{8\pi G}$.
Again, an extra factor of two is present.

It is plausible that this factor of two problem has a good solution.  For
example, although the vector fields (\ref{h10}) obey a standard
$\operatorname{Diff}S^1$ algebra, the corresponding surface deformation
algebra (\ref{h7}) contains three-derivative terms, so the separation of
``central'' and ``boundary'' terms in (\ref{h8}) depends on details of the
choice of $B[\xi]$ (see, for instance, \cite{Catelani}).  The integral over 
time, on the other hand, is more mysterious, although a similar time 
averaging has recently been suggested for the 
$\mathrm{AdS}_2/\mathrm{CFT}_1$ correspondence in a rather
different context \cite{JackiwPi}.  

There are other ways to obtain a dual conformal field theory in the 
(1+1)-dimensional case.  For example, one can use the AdS/CFT 
correspondence to identify the boundary stress-energy tensor and look
at its anomalous transformation properties \cite{Navarro,Larsen,Larsenb}.  
One can also lift the theory to 2+1 dimensions and exploit the 
known properties of the BTZ black hole \cite{Larsen,Larsenb,Navarrob}; 
for the extremal case, this may offer an interesting connection to the 
extremal Kerr/CFT correspondence described in section 
4.ii, 
and might explain the appearance of  a \emph{chiral} CFT 
\cite{Balasubramanian}.  These approaches are designed rather 
particularly for the (1+1)-dimensional black hole, however, and
seem less general than the boundary symmetry method I have focused 
on in this review.

\section{What does this tell us about the states?}
 
One of the chief attractions of the dual conformal field theory approach
to black hole thermodynamics is its universality.  The computations I 
have described depend on only a few minimal features of the black
hole; understanding the asymptotic or near-horizon symmetry is already
enough to determine the entropy.  But this virtue is also a weakness:
it means that the thermodynamics may tell us relatively little about
the true microscopic degrees of freedom of the black hole.

An analogous situation occurs in high energy theory.\footnote{This
analogy was suggested to me by Nemanja Kaloper and John Terning.}  
When a global symmetry is spontaneously broken, Goldstone's 
theorem tells us that there must be a corresponding set of massless 
excitations \cite{Goldstone}.  One such Goldstone boson occurs
for each ``broken'' symmetry generator, and can be viewed as an 
excitation along the ``would-be symmetry'' that is broken by the 
vacuum.  But while Goldstone's theorem gives us some information 
about these massless bosons, it does not tell us how they are 
constructed in terms of the elementary quanta of the theory.

This analogy can be strengthened.  The fundamental symmetry
of general relativity is diffeomorphism invariance.  This is a gauge
symmetry, not a global symmetry of the type directly relevant to
Goldstone bosons.  Still, it remains true that when the
symmetry is broken by the imposition of boundary conditions, 
``would-be pure gauge'' degrees of freedom---excitations that 
would normally be pure gauge, but that do not respect the boundary 
conditions---become physical at the boundary \cite{Carlipl,Carlipk}.  
 
More quantitatively, in Dirac quantization, physical states should be
annihilated by the constraints $H[\xi]$.  But if the central term $K[\xi,\eta]$
in (\ref{b7}) is nonzero, one cannot consistently require such a
condition.  For conformal field theories in two dimensions, this issue 
is well known \cite{Francesco}; one may instead require, for example, 
that only the positive frequency parts of the Virasoro generators 
annihilate physical states.  The result is that certain states that 
would otherwise be considered unphysical must now be allowed.  Since 
the central term is nonzero only at the boundary, these new states are 
naturally associated with the boundary, offering an explanation for 
the fact that black hole entropy depends on area rather than volume.

I cannot say for certain that this is a full explanation for the success
of the dual CFT approach.  There is one case, though, in which the
issue has been investigated more carefully \cite{Carlipj}.  In (2+1)-dimensional
asymptotically anti-de Sitter space, the complete set of boundary terms
for the Einstein-Hilbert action is known explicitly.  These include terms
that are not invariant under diffeomorphisms normal to the boundary.  If
one decomposes the metric near the boundary into a gauge-fixed piece
and a ``pure gauge'' piece, one can determine the induced action for
the ``pure gauge'' portion.  The result is a conformal field theory (a
Liouville theory) with a central charge that exactly matches the expression
(\ref{e13}) for the BTZ black hole.  The relevant states live in the
``nonnormalizable sector'' of the theory, whose quantization is unclear
(though see \cite{Chen}), but this result seems to support the picture of
black hole states as ``would-be pure gauge'' degrees of freedom quite 
nicely.

\section{Conclusion}

In general, black holes are neither two-dimensional nor conformally
invariant.  Nevertheless, we have seen that black hole thermodynamics
often has a simple description in terms of a two-dimensional conformal
field theory at a suitable boundary.  I have focused on the computation
of the most important thermodynamic quantity, the Bekenstein-Hawking
entropy, but similar methods can be employed to analyze greybody factors
and superradiance \cite{Maldacena,Bredberg}.  Similarly, a two-dimensional
conformal field theory approach to \emph{matter} near a black hole
horizon can be used to extract the Hawking temperature and spectrum
\cite{Robinson,Iso,Banerjee,Isob,Bonora}.

While these are important steps, more remains to be done.  A complete
dual description of black hole thermodynamics should also include 
Hawking radiation, which would presumably require a coupling of the
boundary CFT to bulk matter.  For the BTZ black hole, an important step 
in this direction was taken by Emparan and Sachs \cite{Emparan}, who 
considered an external field coupled to the dual CFT and showed that
one could derive the Hawking spectrum from detailed balance.  Their
method can probably be extended to other types of black holes, although
as far as I know this has not yet been done.   

For some important problems in black hole thermodynamics---the information
loss puzzle, for instance \cite{Giddings}, and the question of the final state
of Hawking evaporation---one must go even further, and account for the
back-reaction of Hawking radiation on the black hole.  As noted above, this
probably cannot be accomplished within the framework of a single conformal
field theory: the central charge of the dual CFT typically depends on 
quantities such as the horizon area or angular momentum that change
evaporation.  One can imagine a picture involving a flow from an initial 
CFT to a final one---indeed, in the examples of section \ref{seceg}, the 
direction of such a flow is compatible with Zamolodchikov's $c$ theorem 
\cite{Zam}---but for now, this possibility remains speculative.

\vspace{1.5ex}
\begin{flushleft}
\large\bf Acknowledgments
\end{flushleft}

Portions of this project were carried out at the Peyresq 15 Physics Conference 
with the support of OLAM Association pour la Recherche Fundamentale, 
Bruxelles.  This work was supported in part by Department of Energy grant
DE-FG02-91ER40674.

\end{document}